# Above room temperature multiferroic tunnel junction with the altermagnetic metal CrSb

Long Zhang,[1] Guangxin Ni,[2,3] Junjie He,[4] and Guoying Gao[1,*]

[1]*School of Physics and Wuhan National High Magnetic Field Center, Huazhong University of Science and Technology, Wuhan 430074, P. R. China*

[2]*Department of Physics, Florida State University, Tallahassee, FL 32306, USA*

[3]*National High Magnetic Field Laboratory, Tallahassee, FL 32310, USA*

[4]*Faculty of Science, Charles University, Prague 12843, Czech Republic*

*Contact author: guoying_gao@mail.hust.edu.cn

**ABSTRACT.** Altermagnets with non-relativistic momentum-dependent spin splitting and compensated net magnetic moments have recently garnered significant interest in spintronics, particularly as pinning layers in magnetic tunnel junctions (MTJs). However, room-temperature (RT) altermagnet-based MTJs with tunable tunneling magnetoresistance (TMR) or electroresistance (TER) modulated by multiferroicity remains largely unexplored. Here, we propose an experimentally fabricable above-RT multiferroic MTJ, comprising an altermagnetic metal, ferroelectric barrier, and ferromagnetic metal–epitomized by a $CrSb/In_2Se_3/Fe_3GaTe_2$ heterostructure. Our calculations with first-principles and nonequilibrium Green's function method indicate that the architecture enables magnetically switchable TER, electrically tunable TMR, and dual-mode controllable spin filtering. To disentangle the roles of ferroelectricity and the tunnel barrier, non-ferroelectric $Sb_2Se_3$ and a vacuum gap are exploited as control cases. Remarkably, the system achieves TMR up to 2308 %, TER of 707 %, and near-perfect spin filtering efficiency. Both TMR and TER are considerable for $CrSb/In_2Se_3/Fe_3GaTe_2$ with either Cr or Sb interface. The transport performance is robust under bias voltage. These findings demonstrate the above-RT multiferroic altermagnet-based MTJs and highlight their exciting potential as a versatile platform for next-generation spin dynamics, magnetic-sensing and quantum logic nano-devices.

***Introduction.*** Altermagnets (AMs) [1] are compensated collinear members of the magnetic family, distinct from conventional ferromagnets (FMs), antiferromagnets (AFMs) and ferrimagnets (FiMs) [2-7], they exhibit non-relativistic spin splitting and a vanishing net magnetic moment due to spin and crystalline symmetries [8]. These properties make AMs highly suitable for magnetic memory and logic miniaturized-devices, offering low power consumption, high non-volatility, ultra-fast spin dynamics, and immunity to stray field interference [9-12]. AM combines the superiorities of both FM and AFM, making its entrance into the limelight. Notable examples include CrSb [13,14], MnTe [15,16], $MnF_2$ [1,17], $CaMnO_3$ [18] and $V_2Se_2O$ [19], which are poised to drive revolutionary advances in encoding, storage and transport of information [20].

The current state of the art of the junction non-volatile memory devices is marked by a diversity of approaches, each with distinct characteristics, including ferroelectric (FE) memory [21], magnetic random-access memory [22], and phase-change memory [23]. To attain better device performance, lower cost is needed, with higher performance index (like storage density and capacity), less power consumption, less manufacturing cost (like complexity and yield rate), high reliability (like lifespan and data security), and smaller size and higher integration. The magnetic tunnel junction (MTJ) is a prime design of non-volatile memory [24]. Binary data is encoded by altering the relative magnetization directions of the two magnetic electrodes, aligned magnetic electrodes leads to a low-resistance state while opposed magnetization directions results in a high-resistance state [25,26]. For non-volatile memory, a higher tunneling magnetoresistance (TMR) translates to a more pronounced resistance difference between the two storage states (data bits “0” and “1”), enhancing memory reliability and maintaining the data integrity. The tunneling electroresistance (TER) effect arises in MTJs with a FE barrier, where a significant change in tunnelling resistance occurs due to the reversal of FE polarization. Similar to TMR, a higher TER facilitates a clear distinction between high and low resistance states, improving storage stability and accuracy. Additionally, spin filtering efficiency quantifies a material's ability to selectively transmit electrons based on their spin orientations. A higher spin filtering efficiency sharpens the distinction between different storage states, leading to improved non-volatile memory performance, including enhanced TMR and readout fidelity. Therefore, pursuing large TMR/TER/filtering is of great significance for further advancing the design of high-performance non-volatile memory. For AM-based MTJs, the predicted TMR of Ag/$V_2Te_2O$/BiOCl/$V_2Te_2O$/Ag MTJ reaches 574 %, but it lacks spin filtering and is unregulated [27]. Similarly, a TMR of 150-170 % has been reported for insulating $CoF_2$ and $NiF_2$ with spin filtering [28], but the AMs of $V_2Te_2O$, $CoF_2$ and $NiF_2$ have not yet been experimentally confirmed. Moreover, the crystal-orientation-, interface- and layer-dependent TMR have been proposed in $IrO_2$/$MnF_2$/$CrO_2$ MTJ [29], and $RuO_2$/$TiO_2$/$CrO_2$ MTJ have been shown to detect the Néel vector of $RuO_2$ [30]. However, the Néel temperature ($T_N$) of $MnF_2$ is merely ~67 K [31], the AM in $RuO_2$ faces challenges, with spin splitting potentially arising from lattice distortion and spin-orbit coupling (SOC) [32,33]. In both studies, variations in TMR and spin filtering were attributed to changes in crystal orientation, interface and thickness [29,30], which are intractable to formulate and revert experimentally. Therefore, AMs explorations, particularly their applications, are still in the nascent stage, and there is a pressing need for manipulable room-temperature (RT) AM-based junctions. Additionally, while ferroelectricity and antiferroelectricity could modulate AM [34,35] and the TER has been studied in conventional FM-based MTJs [36,37], there have been no reports on TER in AM-based MTJs.

To address these compelling needs for multiferroic AM-based transport, appropriate electrodes and barriers are pursued. CrSb has recently been spotlighted as a promising altermagnetic metal with a remarkable $T_N$ of ~700 K and momentum-dependent spin splitting, it possesses ferromagnetic Cr-planes and antiferromagnetic couples along the vertical axis [13,38]. Its fully compensated net magnetic moment resists stray magnetic field, which is crucial for stability of spintronic devices. CrSb's non-relativistic momentum-dependent spin splitting means that along the out-of-plane transport direction, some special high-symmetry points are spin-split, while others are spin-degenerate, which allows CrSb to provide a stable and controllable spin-polarized current. Moreover, CrSb combines the advantages of ferromagnetic and antiferromagnetic materials, it has low power consumption and strong non-volatility of ferromagnet, ultrafast spin dynamics and immunity to stray fields of antiferromagnet, making it highly advantageous for spintronic applications. The spin splitting in CrSb mainly comes from exchange coupling, not SOC, and is more robust and easier to exploit in spintronic devices. The electrical inversion with zero magnetic field has been implemented in altermagnetic CrSb when there are vertical components between each pair of spin polarization, Néel vector, and Dzyaloshinskii-Moriya vector [14]. The manipulation of spin splitting in other altermagnets are also reported [34,35,39]. Thus, manipulating spin splitting of CrSb is experimentally achievable. Moreover, $Fe_3GaTe_2$ with a Curie temperature ($T_C$) of 350-380 K and near-half-metal (near-HM) features [40], exhibit strong spin polarization. $In_2Se_3$, a FE semiconductor with a critical temperature of ~700 K [41,42] can serve as a FE-reversible barrier. Considering the ease/possibility of experimental manufacturing of the device, in our designed MTJs, these selected materials are not only experimentally synthesized and readily accessible but also feature above-RT critical temperatures. They all have similar lattices, relatively uncomplicated structures, and good compatibility, all of which facilitate experimental device processing. Thus, CrSb, $Fe_3GaTe_2$ and $In_2Se_3$ are eminently capable for constructing multiferroic tunnel junctions.

In this work, we employ the non-equilibrium Green's function (NEGF) approach combined with density functional theory (DFT) [43,44] to investigate a CrSb/$In_2Se_3$/$Fe_3GaTe_2$ MTJ with magnetically and ferroelectrically tunable TMR, TER and spin filtering. To demonstrate the role of FE, $Sb_2Se_3$ as a topological insulator candidate [45-47], is utilized, and vacuum is introduced to compare barrier effects. Their structures are presented in Figs. 1(a)-1(j). The multistate magneto- and electro-resistances, along with spin filtering in the CrSb/$In_2Se_3$/$Fe_3GaTe_2$ MTJ, offer the potential for revolutionarily multifunctional AM- and FE-based nano-devices.

***Computational details.*** The first-principles calculations for structural relaxation, total energy and electronic structure are performed utilizing the Vienna ab initio Simulation Package (VASP) code based on DFT [48]. The cutoff energy of 600 eV is set, and the convergence criteria for energy and force are chosen as $10^{-6}$ eV and 0.001 eV Å$^{-1}$, respectively. The Γ-centered 12 × 12 × 2 Monkhorst-Pack grid is considered in the Brillouin zone integration, exceptionally, the 12 × 12 × 8 grid is utilized for CrSb bulk. The van der Waals interactions of DFT-D3 method is adopted [49]. A vacuum space of 15 Å is included to eliminate the interlayer interactions from the periodic structures. Perdew-Burke-Ernzerhof (PBE) exchange-correlation functional of the generalized gradient approximation (GGA) with Hubbard U = 0 is carried out with projected augmented wave (PAW) method [44]. The reason for choosing this functional (PBE with U = 0) is presented below.

For CrSb, the electronic structures from calculation results of PBE method with U = 0 yield best agreement with those from experimental data based on angle-resolved photoemission spectroscopy (ARPES) [38,50]. Different U values in bulk $Fe_3GaTe_2$ are tested (see Fig. S1 [51]). As U increases, the bands near the Fermi level in the spin-up channel shift downward, while those in the spin-down channel shift upward. However, a consistent observation is that $Fe_3GaTe_2$ remains its metallic nature across different U values, with a relatively large number of spin-up bands crossing the Fermi level, allowing for relatively high spin polarization. For bulk $Fe_3GaTe_2$, U = 0 extensively presents consistency with experimental results [40,51-53]. Although the magneto-crystalline anisotropy (MCA) calculated by PBE is not as close to the experimental values as the results of local density approximation (LDA), the numerical magnitude of the MCA is not the focus of the present study, while the electronic structure, total energy, and magnetization alignments are the ones that are focused on, and the PBE method with U = 0 for these calculations is basically consistent with the experimental situations including ARPES measurements, which has been widely applied in $Fe_3GaTe_2$ family [40,52,53]. For $In_2Se_3$ and $Sb_2Se_3$, PBE method with U = 0 is commonly used and presents consistency with experimental results [45,54].

The spin transport properties are calculated by the QuantumWise Atomistix ToolKit (ATK) package based on DFT combined with the NEGF method [43]. The 15 × 15 × 150 and 150 × 150 k-meshes are selected for self-consistent calculations and transmission calculations, respectively. The double-z plus polarization (DZP) basis set and the cutoff energy of 150 hartree are used. Similar calculations methods and parameters have been applied in our previous investigations [2,37,55].

***Results and discussion.*** Analogous to the NiAs-type structure [38], CrSb presents the space group of $P6_3/mmc$, where each Cr is surrounded by six Sb atoms [Figs. 1(a) and 1(f)]. This arrangement forms two magnetic sublattices, with the magnetic moments oriented in opposite directions along the *z*-axis, yielding a collinear AFM structure. The band structure of CrSb in both spin-up and spin-down channels doesn't completely overlap along special paths [Figs. 1(k)-1(m)], and the magnetic moments of Cr are ±2.70 $\mu_B$. The high-symmetry points within the Brillouin zone are displayed (see Fig. S2 [51]). CrSb exhibits the *g*-wave-type distribution of spin splitting, with time-reversal ($T$) symmetry combining symmetries beyond spatial translation ($t$) and inversion ($P$) linking the opposite magnetic sublattices, ensuring compensated collinear magnetization that resists disruption from stray fields. The broken $P{\cdot}T$ symmetries in CrSb allow for alternating spin splitting in reciprocal space, stemming from exchange coupling instead of SOC, and forms symmetry-connected spin-momentum locking [1,8]. The above-RT altermagnetic CrSb bulk [13,38] combines preponderances of manipulatable spin splitting properties of FM with the absence of stray field and ultra-fast dynamic response of AFM, making it an invaluable electrode material for antiferromagnetic devices.

$Fe_3GaTe_2$ shares the same space group as CrSb and encompasses five atomic layers, with Te on the outer side and Fe occupying two different positions, one of which is in the central layer with Ga [Figs. 1(b) and 1(g)]. It displays intrinsic above-RT FM [40] and near-HM nature [Fig. 1(n)], with magnetic moments of Fe are 2.11/2.11/1.42 $\mu_B$. One spin channel holds apparently more electronic states than the opposite, engendering high spin polarization and favoring spin filtering effect as a metallic electrode.

For ferroelectrically regulatable TMR and magnetically controllable TER, $In_2Se_3$ is a prominent candidate as a barrier, with a suitable bandgap of 0.67 eV [Fig. 1(o)]. It possesses the space group of $P3m1$ with five atomic layers, with the FE polarization switchable through the motion of the center Se atom [Figs. 1(c), 1(d), 1(h) and 1(i)]. To further explore the effect of FE on transport, non-ferroic $Sb_2Se_3$ and vacuum are used as alternative barriers. $Sb_2Se_3$, a topological insulator candidate, has a near-zero bandgap in bulk form (see Fig. S3 [51]), while its monolayer possesses a bandgap of 0.50 eV [Fig. 1(p)]. The space groups of $Sb_2Se_3$ bulk and monolayer are $R\bar{3}m$ and $P\bar{3}m1$, respectively. $Sb_2Se_3$ monolayer also consists of five atomic layers, alternating between Sb and Se atoms [Figs. 1(e) and 1(j)]. The band structures of all involved materials are consistent with previous reports [38,40,45-47,50,52,56,57], confirming the reliability of our calculations. Apart from the electronic structures, the electric potential distribution in the barrier layer is also examined. Briefly, it's found that the $In_2Se_3$ monolayer possesses a built-in electric field and an intrinsic FE polarization attributable to the vertical asymmetry [Fig. 1(q)]. The potential gradient between In atoms in monolayer is 1.35 eV, with a built-in electric field of 0.31 eV $Å^{-1}$. The potential difference between the two outer surfaces is 1.41 eV, validating the out-of-plane electric polarization. The calculated electrical dipole moment of $In_2Se_3$ monolayer is 0.096 e Å f.u.$^{-1}$, which is close to the reported values (0.094-0.095 e Å f.u.$^{-1}$) [54,58]. In contrast, $Sb_2Se_3$ monolayer shows spatial symmetry, the potential gradient within monolayer and the potential difference between the two surfaces are zero (Fig. 1r), confirming its non-FE nature.

The atomic and orbital behaviors further elucidate the electronic and spin properties of the ground states in the aforementioned materials. To investigate their atomic contributions to the electronic structures near the Fermi level, atom-resolved band structures are demonstrated (see

Figs. S4-S6 [51]). In CrSb, Cr atoms predominantly contribute to the electronic structure. While both spin channels exhibit identical dispersions along the conventional Γ—M—K—Γ—A—L—H—A path, momentum-dependent spin splitting emerges along the special -M'—Γ'—M' and -D—P—D paths. In these cases, the spin-up and spin-down bands are mirror-symmetric with respect to the Γ'/P points, especially near the Fermi level (see Fig. S4 [51]). This spin splitting originates from the oppositely oriented magnetic moments of Cr atomic pairs within the two magnetic sublattices.

In $Fe_3GaTe_2$, high spin polarization is primarily contributed by Fe atoms, while Ga has minimal involvement (see Fig. S5 [51]). For the $In_2Se_3$ and $Sb_2Se_3$ monolayers, Se atoms dominate the electronic states near the valance band maximum and conduction band minimum, underlining their semiconducting character (see Fig. S6 [51]). The intra-orbital-resolved contributions of the dominant atoms are further demonstrated (see Figs. S7-S11 [51]). These orbitals, particularly the *d*-orbitals, $d_{xy}$, $d_{yz}$, $d_{z^2}$, $d_{xz}$ and $d_{x^2-y^2}$ of Cr (Fe) in CrSb ($Fe_3GaTe_2$), and the *p*-orbitals $p_y$, $p_z$ and $p_x$ in $In_2Se_3$ and $Sb_2Se_3$, exhibit distinct contributions near the Fermi level.

In light of the structural and electronic properties, along with their above-RT critical temperatures and experimental feasibility, nano-MTJs consist of CrSb, $Fe_3GaTe_2$, $In_2Se_3$ and $Sb_2Se_3$ layers are constructed, as shown in Figs. 2(a)-2(d). The CrSb/$In_2Se_3$/$Fe_3GaTe_2$ and CrSb/$Sb_2Se_3$/$Fe_3GaTe_2$ MTJs are fully relaxed. Regarding the CrSb/vacuum/$Fe_3GaTe_2$ MTJ, the vacuum layer thickness ($d_V$) in this MTJ was deliberately chosen, and similar man-made vacuum layers have been utilized in prior studies [59-61]. The lattice parameters of CrSb, $Fe_3GaTe_2$, $In_2Se_3$ and $Sb_2Se_3$ are 4.147, 4.030, 4.075 and 4.055 Å, respectively. Consequently, the lattice mismatches for the $CrSb/In_2Se_3/Fe_3GaTe_2$, $CrSb/Sb_2Se_3/Fe_3GaTe_2$, and $CrSb/vacuum/Fe_3GaTe_2$ MTJs are within 1.52 %, 1.68 % and 1.45 %, respectively. Experimentally, lattice mismatch ratios can reach 5 % in Fe/MgO/Fe MTJ [62]; ~7 % in graphene-contacted $GaSe/In_2Se_3$ junction [63]; 4.3 % in $Co_2MnSi/MgO/Co_2MnSi$ MTJ [64]; and 6.5 % in $Fe/MgAlO_x/Fe_4N$ MTJ [65]. These precedents suggest that our present mismatch of 1.45-1.68 % is within the experimentally feasible range. We define parallel ($M_{\uparrow\uparrow}$) and antiparallel ($M_{\uparrow\downarrow}$) magnetization configurations as the interlayer ferromagnetic and antiferromagnetic couplings of $Fe_3GaTe_2$, respectively, and denote two opposite directions of FE polarization directions of the $In_2Se_3$ barrier as $P_\downarrow$ and $P_\uparrow$. Van der Waals interactions occur between $In_2Se_3/Sb_2Se_3$ and $Fe_3GaTe_2$, while CrSb bond directly to $In_2Se_3/Sb_2Se_3$ by virtue of interlayer spacing.

To quantitively compare magnetic and FE tunneling phenomena across these MTJs, we introduce the following key metrics. Specifically, the spin filtering efficiency $\eta$ can be defined as

$$\eta = \left| \frac{T_\uparrow - T_\downarrow}{T_\uparrow + T_\downarrow} \right| \times 100\% \tag{1}$$

where $T_\uparrow$ and $T_\downarrow$ are the transmission coefficients at the Fermi level for spin-up and spin-down channels, respectively. TMR can be obtained by

$$\text{TMR} = \frac{T_\text{P} - T_\text{AP}}{T_\text{AP}} \times 100\% \tag{2}$$

with $T_\text{P}$ and $T_\text{AP}$ representing transmission coefficients at the Fermi level for parallel and antiparallel magnetization alignments, respectively. Similarly, TER can be attained from

$$\text{TER} = \frac{\left| T_\uparrow - T_\downarrow \right|}{\min\left(T_\uparrow, T_\downarrow\right)} \times 100\% \tag{3}$$

in which $T_\uparrow$ and $T_\downarrow$ are the transmission coefficients at the Fermi level for the opposite FE polarization directions of the barrier layer.

Transmission is further resolved by in-plane wave vector $\vec{k}_{//} = (k_x, k_y)$, which is perpendicular to the transport direction. The transmission coefficient $T_P$ and $T_{AP}$ can be obtained by $T_P = \sum_{\vec{k}_{//}} T_P(\vec{k}_{//}) / N_k$ and $T_{AP} = \sum_{\vec{k}_{//}} T_{AP}(\vec{k}_{//}) / N_k$, respectively, in which the number of $k$-points is represented by $N_k$. The full spin- and $\vec{k}_{//}$-resolved transmission spectra across the MTJ of $CrSb/In_2Se_3/Fe_3GaTe_2$ in energy and momentum space are illustrated in Figs. 3(a)-3(r). Transmission is concentrated near the M and K points, with minimal contributions around Γ and other points. For $CrSb/In_2Se_3/Fe_3GaTe_2$ with $P_\downarrow$ and $M_{\uparrow\uparrow}$, the spin-up channel's transmission coefficient at the Fermi level significantly exceeds that of the spin-down channel [Figs. 3(c) and 3(d)], enabling efficient tunneling in the spin-up channel and yielding a near-perfect spin filtering efficiency ($\eta$ = ~100 %). Switching the magnetization of $Fe_3GaTe_2$ to antiparallel ($M_{\uparrow\downarrow}$) suppresses tunneling in both spin channels, with a slightly higher spin-down coefficient [Figs. 3(g) and 3(h)], resulting in a moderate $\eta$ of 45 %. For $P_\uparrow$, pronounced tunneling and high spin filtering ($\eta$ = 90 % for $M_{\uparrow\uparrow}$, 86 % for $M_{\uparrow\downarrow}$) occur, though spin-resolved behaviors differ: $T_\uparrow$ dominates in $M_{\uparrow\uparrow}$, while $T_\downarrow$ prevails in $M_{\uparrow\downarrow}$, achieving spin filtering in both channels. As summarized in Table I, TMR reaches 1031 % ($P_\downarrow$) and 132 % ($P_\uparrow$), while TER is 14 % ($M_{\uparrow\uparrow}$) and 328 % ($M_{\uparrow\downarrow}$). Thus, the FE polarization direction of $In_2Se_3$ barrier strongly influences TMR, while $Fe_3GaTe_2$ magnetization configuration tunes TER, with both modulating spin filtering effect. Notably, it's essential to design devices with feasible smaller lattice mismatch rates and higher TER ratios. We here give a comparison with previous works. The TER of 328 % in our AM-based multiferroic MTJ with an $In_2Se_3$ monolayer is higher than those of 62 % in $Fe_mGeTe_2/In_2Se_3/Fe_nGeTe_2$ (m, n = 3, 4, 5; m ≠ n) [66] and 90 %-270 % in others [67,68] , but two of them feature smaller mismatches of 1 % [66] and 0.3 % [67] than our reported 1.45-1.68 %. Additionally, the mismatch is less than 2 % and the TER is 52 % in $SrRuO_3/BaTiO_3/SrRuO_3$ [69], and the mismatch is about 5 % and the TER is 30 % in $TiTe_2/Mn_2Se_3/In_2S_3/Mn_2Se_3/TiTe_2$ with $In_2S_3$ monolayer [70]. Both mismatches are larger and both TERs are smaller compared with our present work. The mismatch is 5.6 % and the TER is 349 % in $Fe_3GeTe_2/MnSe/Fe_3GeTe_2$ [71], the TER is slightly larger but the mismatch is quite larger. These comparisons indicate the competitive TER in $CrSb/In_2Se_3/Fe_3GaTe_2$.

To assess the FE effect, we examined tunneling transport through the non-FE $Sb_2Se_3$ barrier in $CrSb/Sb_2Se_3/Fe_3GaTe_2$. For $M_{\uparrow\uparrow}$, negligible transmission occurs in the spin-down state, but not in the spin-up one, while for $M_{\uparrow\downarrow}$, both channels show comparable coefficients, yielding a TMR of 198 %. Spin filtering is near-perfect ($\eta$ = ~100 %) for $M_{\uparrow\uparrow}$ and moderate ($\eta$ = 62 %) for $M_{\uparrow\downarrow}$ (Table II). Similarly, a vacuum barrier in $CrSb/vacuum/Fe_3GaTe_2$ permits spin-up tunneling for $M_{\uparrow\uparrow}$ but blocks both channels for $M_{\uparrow\downarrow}$, achieving a high TMR of 2308 %, and a spin filtering of ~100 % ($M_{\uparrow\uparrow}$) and moderate ($M_{\uparrow\downarrow}$) of the $Fe_3GaTe_2$ electrodes. The TMR across $CrSb/In_2Se_3/Fe_3GaTe_2$ with $P_\downarrow$ is 1031 %, while it reaches 2308 % across $CrSb/vacuum/Fe_3GaTe_2$, a difference that must be clearly highlighted. The larger change of $T_\downarrow$ with $M_{\uparrow\downarrow}$ compared to the change in $T_\uparrow$ with $M_{\uparrow\uparrow}$ results in a relatively higher TMR when using a vacuum (Figs. 3(a) and 3(b), and Tables I and II). Although transmissions are concentrated around the M and K points, there is a relatively large difference between $In_2Se_3$ [Fig. 3(h)] and vacuum [Fig. 3(r)] in the spin-down channel with $M_{\uparrow\downarrow}$, while the differences are relatively minor in spin-up channel with $M_{\uparrow\uparrow}$ [Figs. 3(c) and 3(o)].

To understand the underlying mechanisms behind the observed regulable and multiplicate TMR, TER and spin filtering across these MTJs, we analyzed the spin-resolved local density of states (LDOS), and spin- and layer-resolved projected device density of states (PDDOS) (Figs. 4(a)-4(l) and see Figs. S12(a)-S12(l) [51]), where the source of contribution is set as the left CrSb electrode. The spin polarization of non-magnetic barrier layers chiefly stems from the magnetic proximity effect of CrSb and $Fe_3GaTe_2$. In CrSb/$In_2Se_3$/$Fe_3GaTe_2$ MTJ with $M_{\uparrow\uparrow}$, states near the Fermi level are predominately spin-up in both FE polarization states, with negligible spin-down states [Figs. 4(a)-4(f)]. This leads to strong spin filtering effects. Since total electronic states remain similar between $P_\uparrow$ and $P_\downarrow$, TER is minimal in this configuration. For the $M_{\uparrow\downarrow}$ configuration presented in Figs. 4(g)-4(l), tunneling is generally suppressed. However, spin-down states prevail, resulting in spin-down filtering, particularly with $P_\uparrow$ polarization, where more available states in $Fe_3GaTe_2$ enhance TER. Considering magnetic tunneling, under the circumstance with $P_\downarrow$ of $In_2Se_3$ barrier, more black (low-density) regions in both spin channels for the $M_{\uparrow\downarrow}$ configuration, and in the spin-down channel for the $M_{\uparrow\uparrow}$ configuration, compared to the spin-up channel for $M_{\uparrow\uparrow}$ near the Fermi level [Figs. 4(a), 4(b), 4(g), and 4(h)]. This imbalance in available electronic states between $M_{\uparrow\uparrow}$ and $M_{\uparrow\downarrow}$ leads to a substantial difference in tunneling conductance, giving rise to a high TMR. However, when the polarization is switched to $P_\uparrow$, the summed electronic states near the Fermi level for $M_{\uparrow\uparrow}$ and $M_{\uparrow\downarrow}$ become nearly indistinguishable, thus weakening the conductance contrast and resulting in a much lower TMR. Notably, with $M_{\uparrow\uparrow}$, the magnitude of the density of states near the Fermi level in the spin-up channel through the right $Fe_3GaTe_2$ layers is slightly higher for CrSb/$In_2Se_3$/$Fe_3GaTe_2$ with $P_\downarrow$ than for CrSb/vacuum/$Fe_3GaTe_2$ (Figs. 4(a)-4(c) and see Figs. S12(g)-S12(h) [51]), leading to the slightly higher transmission coefficients for CrSb/$In_2Se_3$/$Fe_3GaTe_2$ with $P_\downarrow$. However, with $M_{\uparrow\downarrow}$, the difference in the density of states in the spin-down channel is striking, CrSb/$In_2Se_3$/$Fe_3GaTe_2$ with $P_\downarrow$ exhibits a large number of electronic states near the Fermi level [Figs. 4(g)-4(i)], while CrSb/vacuum/$Fe_3GaTe_2$ shows only a small spike with relatively few electronic states (see Figs. S12(j)-S12(l) [51]). This yields a significant different transmission in the spin-down channel with $M_{\uparrow\downarrow}$. These differences in the density of states in the main contributing channels at $M_{\uparrow\uparrow}$ and $M_{\uparrow\downarrow}$ lead to markedly different TMRs. This suggests that the $In_2Se_3$ and vacuum intermediate layers noticeably affect the electronic structures of CrSb and $Fe_3GaTe_2$ due to magnetic proximity effects, which in turn cause substantial changes in the transport properties. The altermagnetic electronic structures of CrSb, particularly when it is terminated as a lead, can be affected. As the lead, CrSb acts as a semi-infinite electrode material, resembling a bulk rather than a few layers. It is in contact with the barrier layer. The PDDOS in the MTJs (Figs. 4(c), 4(f), 4(i), and 4(l) and see Figs. S12(c), S12(f), S12(i), and S12(l) [51]) indicate that the electronic structures of the unit-cell CrSb at and near the terminal are impacted, but those of the internal unit-cell CrSb far from the terminal may remain unaffected. For example, in the CrSb/$In_2Se_3$/$Fe_3GaTe_2$ MTJ with Cr interface and $M_{\uparrow\uparrow}$-$P_{\downarrow}$, these five sub-figures in Fig. 4(c) display the calculated electronic structures (PDDOS) of unit-cell CrSb, CrSb, $In_2Se_3$, $Fe_3GaTe_2$, and $Fe_3GaTe_2$ in the central scattering region in sequence. It can be observed that the electronic states in the two spin channels of the terminated unit-cell CrSb close to $In_2Se_3$ exhibit significant differences (the second sub-figure in Fig. 4(c)), indicating that the altermagnetic electronic structure at the terminal is altered. In contrast, for the unit-cell CrSb not at the terminal and farther from $In_2Se_3$, the difference in electronic states between the two spin channels is negligible (the first sub-figure in Fig. 4(c)), suggesting that it largely retains its altermagnetic electronic structure. At a greater distance from the terminal, the electronic structure of the internal unit-cell CrSb approaches that of the unit-cell bulk. Therefore, the unit-cell CrSb far from the terminal remains unaffected, and should retain the altermagnetic electronic structure. Besides that, the corresponding LDOS and PDDOS features for the single-ferroic CrSb/$Sb_2Se_3$/$Fe_3GaTe_2$ and CrSb/vacuum/$Fe_3GaTe_2$ MTJs display certainly similar characteristic and are therefore not discussed in detail.

To more intuitively illustrate the distribution of electronic states and transport characteristics within the constructed MTJs during operation, the tunneling transport pathways are explicitly depicted (Figs. 5(a)-5(h) and see Figs. S13(a)-S13(h) [51]), in which the left CrSb electrodes serving as the electron source. For the $M_{\uparrow\uparrow}$ of $Fe_3GaTe_2$, the number of spin-up transmission pathways is significantly greater than that of spin-down, under both $P_{\downarrow}$ and $P_{\uparrow}$ of $In_2Se_3$ [Figs. 5(a)-5(d)]. In contrast, for $M_{\uparrow\downarrow}$-$P_{\uparrow}$, the spin-down pathways dominate, while spin-up channels are obviously suppressed [Figs. 5(g) and 5(h)], resulting in a high spin filtering effect. However, in the $M_{\uparrow\downarrow}$-$P_{\downarrow}$ case [Figs. 5(e) and 5(f)], the numbers of spin-up and spin-down become comparable, leading to a relatively weak spin filtering efficiency. Interestingly, the spin-down channels under $M_{\uparrow\downarrow}$-$P_{\downarrow}$ exhibit more transport pathways than the spin-down channels under $M_{\uparrow\uparrow}$-$P_{\downarrow}$, but still fewer than the spin-up channels in $M_{\uparrow\uparrow}$-$P_{\uparrow}$, situating them in an intermediate position. Under $P_{\downarrow}$, the $M_{\uparrow\uparrow}$ configuration exhibits the largest number of transport pathways [Figs. 5(a), 5(b), 5(e) and 5(f)], particularly in the spin-up channel, composing a large contrast compared to the $M_{\uparrow\downarrow}$ and engendering an eminent TER. What's more, under $P_{\uparrow}$, the spin-up channel in $M_{\uparrow\uparrow}$ and the spin-down channel in $M_{\uparrow\downarrow}$ dominate [Figs. 5(c), 5(d), 5(g) and 5(h)], and since these are roughly comparable, the resulting TMR is much lower. Thus, the high (low) TER can be interpreted by the dominant pathways in spin-down (-up) channel under $M_{\uparrow\downarrow}$ ($M_{\uparrow\uparrow}$) configurations, modulated by the direction of FE polarization. These results from transport pathway distributions are fully consistent with the prior analyses of transmission coefficients, LDOSs and PPDOSs discussed above. In the spin-up channel with $M_{\uparrow\uparrow}$, CrSb/$In_2Se_3$/$Fe_3GaTe_2$ with $P_{\downarrow}$ exhibits more transmission pathways than $CrSb/vacuum/Fe_3GaTe_2$ (Figs. 5(a) and see Fig. S13(e) [51]), but in the spin-down channel with $M_{\uparrow\downarrow}$, the difference is even more pronounced, with $CrSb/In_2Se_3/Fe_3GaTe_2$ with $P_\downarrow$ showing a clearer excess of pathways compared to $CrSb/vacuum/Fe_3GaTe_2$ (Figs. 5(f) and see Fig. S13(h) [51]). This differential magnification of electron pathways results in a lower TMR of $CrSb/In_2Se_3/Fe_3GaTe_2$ with $P_\downarrow$ compared to $CrSb/vacuum/Fe_3GaTe_2$. As for the single-ferroic $CrSb/Sb_2Se_3/Fe_3GaTe_2$ and $CrSb/vacuum/Fe_3GaTe_2$ MTJs (see Fig. S13 [51]), the tunneling pathway characteristics largely mirror those of the $CrSb/In_2Se_3/Fe_3GaTe_2$ MTJ shown in Fig. 5, and are therefore not elaborated upon here.

The interface configuration is crucial for performance prediction and further experimental realization. In addition to the above transport performance based on the devices with Cr interface in Figs 2(a)-2(d), we further consider another interface Sb and calculate the transport performance. The devices with Sb interface are illustrated (see Fig. S14 [51]), and the spin-dependent transmission coefficients as a function of energy are presented (see Fig. S15 [51]). Spin-dependent electron transmission $T_\uparrow$ and $T_\downarrow$, TMR, TER, and spin filtering efficiency $\eta$ are listed in Tables III and IV. The spin- and layer-resolved PDDOS under different magnetic and FE configurations are displayed in Figs. 6(a)-6(h).

The TMR and spin filtering efficiency across $CrSb/In_2Se_3/Fe_3GaTe_2$ MTJ show slight differences between the Cr interface [Figs. 2(a) and 2(b)] and Sb interface (see Figs. S14(a) and S14(b) [51]). With $P_\downarrow$, the TMR is 1056 % with Sb interface and that is 1031 % with Cr interface. The spin filtering efficiency is near-perfect (~100 %) with $M_{\uparrow\uparrow}$ while it is moderate (~50 %) with $M_{\uparrow\downarrow}$ for both interfaces. With $P_\uparrow$, the TMRs are 136 % and 132 % with Sb and Cr interfaces, respectively, and the spin filtering efficiencies are all high (~90 %). The transport performance with Sb interfaces will be discussed in the following as a comparison with the Cr interfaces. As shown in Fig. S15 [51], with $P_{\downarrow}$, the electrons in the spin-up channel of $M_{\uparrow\uparrow}$ dominate transmission, while the transmission in spin-down channel is minimal, causing a near-perfect spin filtering. The transmissions in spin-up and spin-down channels are comparable with $P_{\downarrow}$ and $M_{\uparrow\downarrow}$, yielding a moderate spin filtering. The transmission in the spin-up channel of $M_{\uparrow\uparrow}$ significantly exceeds that in the spin-down channel of $M_{\uparrow\uparrow}$ and those in the both spin channels of $M_{\uparrow\downarrow}$, resulting in a high TMR with $P_{\downarrow}$. There is a peak of electronic states in spin-up channel around the Fermi level across the right $Fe_3GaTe_2$ layers in the device with $P_{\downarrow}$ and $M_{\uparrow\uparrow}$ [Fig. 6(a)], but the spin-down channel doesn't have, facilitating electrons transmission in the spin-up channel. For $P_{\downarrow}$ and $M_{\uparrow\downarrow}$, there are some states near the Fermi level in the spin-down channel [Fig. 6(c)], making electron transmission relatively easier in this channel. The electronic states near the Fermi level are fewer for $M_{\uparrow\downarrow}$ than for $M_{\uparrow\uparrow}$ [Figs. 6(a) and 6(c)], contributing to a high TMR with $P_{\downarrow}$. With $P_{\uparrow}$, transmission is dominated by the spin-up channel in $M_{\uparrow\uparrow}$ and by the spin-down channel in $M_{\uparrow\downarrow}$, both producing high spin filtering effects. The transmission coefficients are larger in $M_{\uparrow\uparrow}$ than in $M_{\uparrow\downarrow}$, but the difference is not substantial, and the electronic states support this conclusion [Figs. 6(b) and 6(d)]. This leads to an inferior TMR with $P_{\uparrow}$. It's evident that devices with Sb and Cr interfaces of CrSb clearly exhibit different distributions of electronic states near the Fermi level [Figs. 6(a)-6(d) and Figs. 4(c), 4(f), 4(i) and 4(l)], the left CrSb electrodes serve as the source, leading to a different number of electronic states crossing the barrier to the right side. However, our main conclusions for this part remain unaffected. What's more, the TERs using Sb interface (65 % for $M_{\uparrow\uparrow}$ and 707 % for $M_{\uparrow\downarrow}$) are higher than those using Cr interface (14 % for $M_{\uparrow\uparrow}$ and 328 % for $M_{\uparrow\downarrow}$), but the trend is similar: relatively low TER for $M_{\uparrow\uparrow}$ and high TER for $M_{\uparrow\downarrow}$. With $M_{\uparrow\uparrow}$, there are a certain amount of electronic states around the Fermi level for both $P_{\downarrow}$ and $P_{\uparrow}$ [Figs. 6(a) and 6(b)], leading to similar transmissions, especially in the dominant spin-up channel, so this results in inferior TER. While with $M_{\uparrow\downarrow}$, the transmission in the spin-down channel of $P_{\uparrow}$ obviously surpasses that in the spin-up channel of $P_{\uparrow}$ and in both spin channels of $P_{\downarrow}$, the electronic states in the spin-down channel of $P_{\uparrow}$ are predominant [Figs. 6(c) and 6(d)], generating a high TER. For $M_{\uparrow\downarrow}$, the higher TER of 707 % with Sb interface is mainly attributed to more drastically reduced transmission coefficients in both spin channels of $P_{\downarrow}$ compared to that of 328 % with Cr interface. The density of states near the Fermi level through the right $Fe_3GaTe_2$ layers is significantly reduced from Cr interface to Sb interface [Figs. 4(i) and 6(c)].

For CrSb/$Sb_2Se_3$/$Fe_3GaTe_2$ and CrSb/vacuum/$Fe_3GaTe_2$ MTJs, the TMR exhibits significant changes, while the spin filtering effect remains similar with both Cr and Sb interfaces. Near-perfect spin filtering efficiency is achieved with $M_{\uparrow\uparrow}$, regardless of whether Sb or Cr interfaces are employed, whereas the spin filtering efficiency is comparatively low with $M_{\uparrow\downarrow}$. In the case of the $Sb_2Se_3$ barrier, the TMR increases to 325 % for Sb interface from 198 % for Cr interface. This difference is primarily attributed to a substantial reduction in the transmission coefficient in the spin-up channel with $M_{\uparrow\downarrow}$. The density of states reveals a consistent pattern: due to the distinct electronic states source from the left CrSb, the electronic states penetrating through the right side are minimal in the spin-up channel near the Fermi level (see Figs. S12(f) [51], and Fig. 6(f)). Conversely, the transmission coefficients in the spin-down channel with $M_{\uparrow\downarrow}$ and in both spin channels with $M_{\uparrow\uparrow}$ show little variation, with their electronic states near the Fermi level being comparable (see Figs. S12(c) and S12(f) [51], and Figs. 6(e) and 6(f)). This results in a relatively high TMR across CrSb/$Sb_2Se_3$/$Fe_3GaTe_2$ MTJ with Sb interface, but the difference isn't substantial. For the vacuum barrier, the TMR undergoes an obvious change since the transmission coefficients in the dominant spin-up channel with $M_{\uparrow\uparrow}$ are not as large when using Sb interface as they are using Cr interface. The differential electronic states generated by the left CrSb source with different interfaces lead to a much smaller peak in the density of states near the Fermi level in the spin-up channel through the right $Fe_3GaTe_2$ in devices with Sb interface compared to those with Cr interface (Figs. 6(g) and see Fig. S12(i) [51]). The spin-down channel exhibits the opposite trend, but the increment is not significant enough to offset the overall reduction in the transmission coefficients with $M_{\uparrow\uparrow}$. The electronic states in both spin channels near the Fermi level remain largely unchanged with $M_{\uparrow\downarrow}$ (Figs. 6(h) and see Fig. S12(l) [51]), resulting in minimal changes in the transmission coefficients and, consequently, a reduced TMR. To conclude, the spin filtering efficiencies of all devices are robust against interface variations; the most notable changes are observed in the TMR across the CrSb/vacuum/$Fe_3GaTe_2$ MTJ; the TMR across CrSb/$In_2Se_3$/$Fe_3GaTe_2$ MTJ shows little variation; and the TER across CrSb/$In_2Se_3$/$Fe_3GaTe_2$ and the TMR across CrSb/$Sb_2Se_3$/$Fe_3GaTe_2$ increase from Cr interface to Sb interface.

The interface effect on transport performance is evident, yet the stability in energy is equally critical for experimental realization. Our calculated total energies of the devices with Sb interface (see Figs. S14(a)-S14(d) [51]) are substantially higher by 1.387, 1.325, 1.883 and 0.952 eV, respectively, than those of devices with Cr interface [Figs. 2(a)-2(d)]. This indicates that the devices with Cr interface are more stable and thus more experimentally feasible. Consequently, the transmission performance of devices with Cr interface is prioritized. Nevertheless, the transport properties of devices with Sb interface are still valuable and can serve as a reference.

The choice of interface notably influences the numerical and quantitative transport results, but its effect on the qualitative results is not relatively large. The devices with Cr interface demonstrate better stability compared to those with Sb interface. Key characteristics such as the magnetically switchable TER, electrically tunable TMR, and dual-mode controllable spin filtering, remain robust across different interfaces. These findings suggest that our devices have the potential for experimental realization, as they exhibit excellent transport performance alongside acceptable interface compatibility.

Furthermore, the effect of bias voltage on the MTJ is also crucial. Taking the $CrSb/In_2Se_3/Fe_3GaTe_2$ MTJ with Cr interface as a representative, the bias voltage effect on this MTJ as shown in Figs. 7(a)-7(l), compared with the equilibrium state (Table I). It is found that, with $M_{\uparrow\uparrow}$, as the bias voltage increases, the dominant current in spin-up channel increases significantly, while the current in spin-down channel also rises but remains at a relatively low level (Figs. 7(a) and 7(b)). This leads to a high spin filtering efficiency for both $P_{\downarrow}$ and $P_{\uparrow}$ with $M_{\uparrow\uparrow}$ (Figs. 7(e) and 7(f)). The transport performance under equilibrium and bias exhibits similarities, with the spin-up channel being the dominant one and high spin filtering efficiency being achievable. Additionally, the TER is maintained at a very low level (~10 %) at equilibrium and bias conditions (Fig. 7(k) and Table I). For $M_{\uparrow\uparrow}$ and $P_{\downarrow}$, the spin filtering efficiency decreases with increasing bias voltage (Fig. 7(e)), whereas for $M_{\uparrow\uparrow}$ and $P_{\uparrow}$, the spin filtering efficiency remains relatively stable at a high level (~90 %) even as the bias voltage increases (Fig. 7(f)). With $M_{\uparrow\downarrow}$, the spin-down channel dominates the transport, both in equilibrium and under bias conditions (Figs. 7(c) and 7(d), and Table I). As the bias voltage increases, the spin-down current rises significantly, while the spin-up current remains relatively low. With $M_{\uparrow\downarrow}$ and $P_{\downarrow}$, the spin filtering efficiency at small bias is close to the value at the equilibrium state (a medium value), but increases with higher biases due to the rise in the spin-down current, thereby enhancing the spin filtering effect (Figs. 7(c) and 7(g)). With $M_{\uparrow\downarrow}$ and $P_{\uparrow}$, the spin filtering efficiency at small and no biases stays high value (~80 %) (Fig. 7(h) and Table I), with minor fluctuations as the bias voltage increases, attributed to differences in the two spin channels' currents (Fig. 7(h)). Furthermore, with $M_{\uparrow\downarrow}$, the TER increases with the increase of bias voltage (Fig. 7(l)), which is owing to the difference in current under the two ferroelectric polarization conditions. The maximum TER under bias voltage can reach ~600 % at bias of 175 mV. Regarding TMR, with $P_{\downarrow}$, it decreases as the bias voltage increases (Fig. 7(i)), as the current difference between $M_{\uparrow\uparrow}$ and $M_{\uparrow\downarrow}$ cases is reduced. At zero bias voltage, the TMR is 1031 %, and it is only ~50 % at 200 mV (Fig. 7(i) and Table I). Conversely, with $P_{\uparrow}$, the TMR generally increases with bias voltage, rising from 132 % at zero bias to a maximum of ~1000 % around 175 mV (Fig. 7(j) and Table I). Overall, under bias voltage, high TMR and TER can still be achieved by the ferroelectric and magnetic configurations, respectively, further highlighting the applications of multiferroic tunnel junctions.

The tunneling transport mechanisms of altermagnetic/non-magnetic/(anti)ferromagnetic MTJs can be schematized as Figs. 5(i) and 5(j). Unlike conventional MTJs that require localized magnetic fields to switch the magnetization of the free layer while keeping the pinning layer fixed, our AM-based MTJs allow for a broader magnetic-field operation range (through the whole MTJ), which is beneficial for minimizing the spin device. This is because CrSb, as an AM, possesses macroscopic antiferromagnetic order, and reversing the orientations of its intrinsic opposing magnetic moments doesn't affect overall transport performance. As such, CrSb is exceptionally well-suited as a pinning layer in our proposed MTJs. These multi-physical field couplings endow our MTJs with unique electron and spin transport properties and modulation mechanisms. Furthermore, the schematics of our multiferroic and single-ferroic tunneling junctions are illustrated in Figs. 2(e) and 2(f), these junctions exhibit multiple resistance states and spin filtering effects, which can be reversibly tuned through switches of magnetism and ferroelectricity, unveiling a striking horizon of multifunctional nano-devices in spintronics and electronics.

***Conclusion.*** To conclude, using a combined NEGF and DFT framework, we have proposed multiferroic CrSb/$In_2Se_3$/$Fe_3GaTe_2$ and single-ferroic CrSb/$Sb_2Se_3$/$Fe_3GaTe_2$ and CrSb/vacuum/$Fe_3GaTe$ MTJs to systematically investigate magneto- and FE-switchable spin and charge transport behaviors. Notably, the CrSb/$In_2Se_3$/$Fe_3GaTe_2$ MTJ exhibits a remarkable TMR of 1031 %, a TER of 707 %, and near-perfect spin filtering. A vacuum barrier also yields a high TMR of 2308 %. The interlayer magnetic alignment of $Fe_3GaTe_2$ controls the electroresistance states, while the FE polarization of $In_2Se_3$ modulates magnetoresistance, and both jointly regulate spin filtering efficiency. The interfacial effect (Cr and Sb interfaces) on the spin transport properties is also explored, and both interfaces exhibit excellent transport performance. The bias voltage effect on our MTJs highlights the performance robustness. These multistate, tunable effects in experimentally feasible and above-RT CrSb/$In_2Se_3$/$Fe_3GaTe_2$ MTJs represent an illuminating avenue for the development of AM-based multiferroicity-tunable spintronic and electronic nano-devices, with robust, multifunctional memory and logic capabilities.

***Acknowledgments.*** We sincerely thank Prof. Junwei Liu from Hong Kong University of Science and Technology, Prof. Tian Qian from Institute of Physics, Chinese Academy of Sciences, and Prof. Menghao Wu from Huazhong University of Science and Technology, for their beneficial discussion about altermagnetic features and transport properties. Guoying Gao acknowledges support from the National Natural Science Foundation of China (Grant No. 12174127). Guangxin Ni acknowledges support from U.S. Department of Energy (DE-SC0022022), U.S. National Science Foundation (DMR-2145074), and U.S. American Chemical Society Doctoral New Investigator (PRF# 66465-DNI10) Grants Program.

***Data availability.*** The data that support the findings of this article are not publicly available upon publication because it is not technically feasible and/or the cost of preparing, depositing, and hosting the data would be prohibitive within the terms of this research project. The data are available from the authors upon reasonable request.

## Figures and Tables

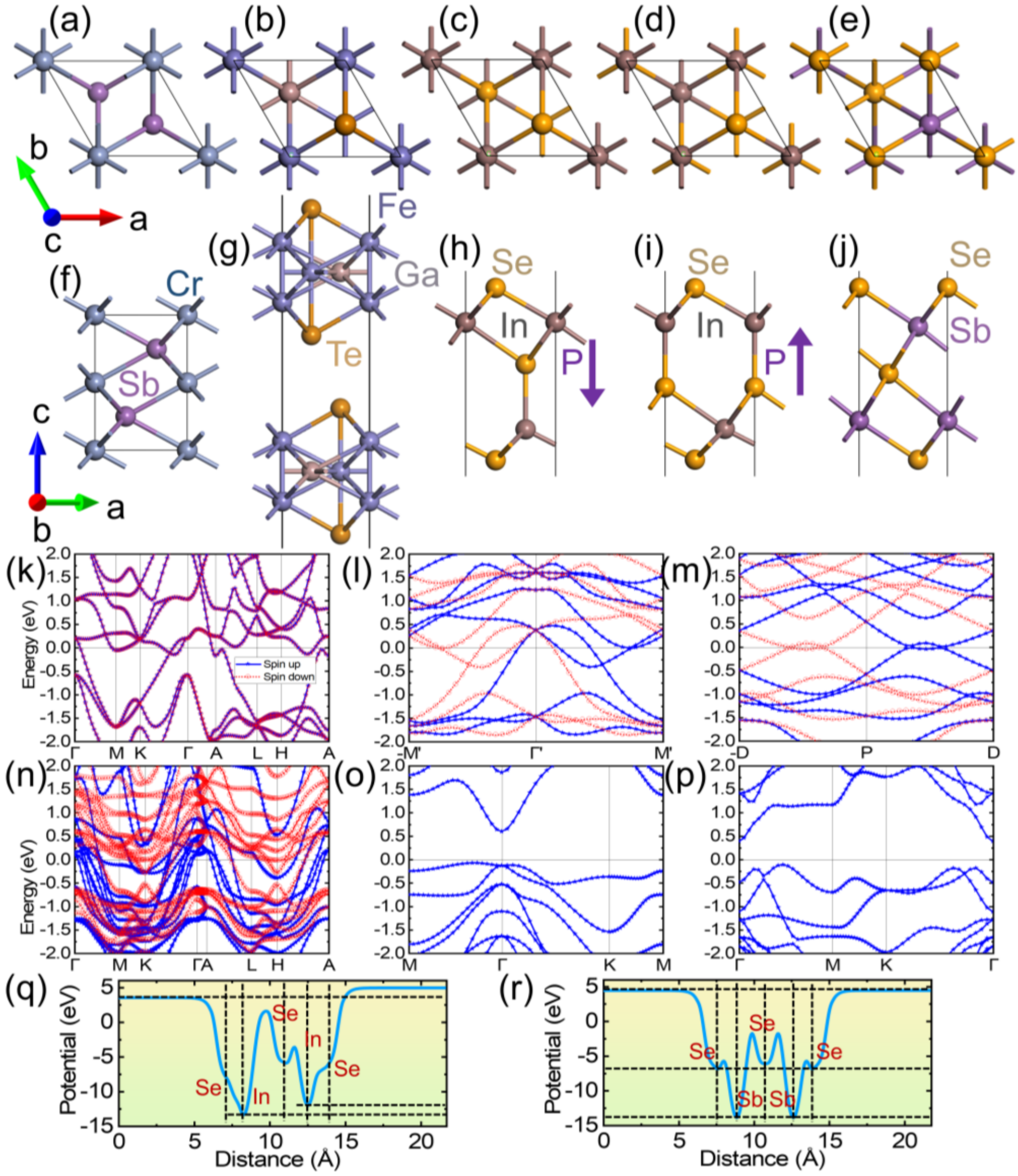

FIG. 1. Top (a-e) and side (f-j) views for the crystal structures of CrSb bulk, $Fe_3GaTe_2$ bulk, $In_2Se_3$ monolayers with opposite ferroelectric polarization states, and $Sb_2Se_3$ monolayer. Band structures of CrSb bulk along different high-symmetry paths (k-m), $Fe_3GaTe_2$ bulk (n), $In_2Se_3$ monolayer (o), and $Sb_2Se_3$ monolayer (p). Averaged electrostatic potential profiles along the out-of-plane *c*-axis direction for $In_2Se_3$ monolayer (q) and $Sb_2Se_3$ monolayer (r).

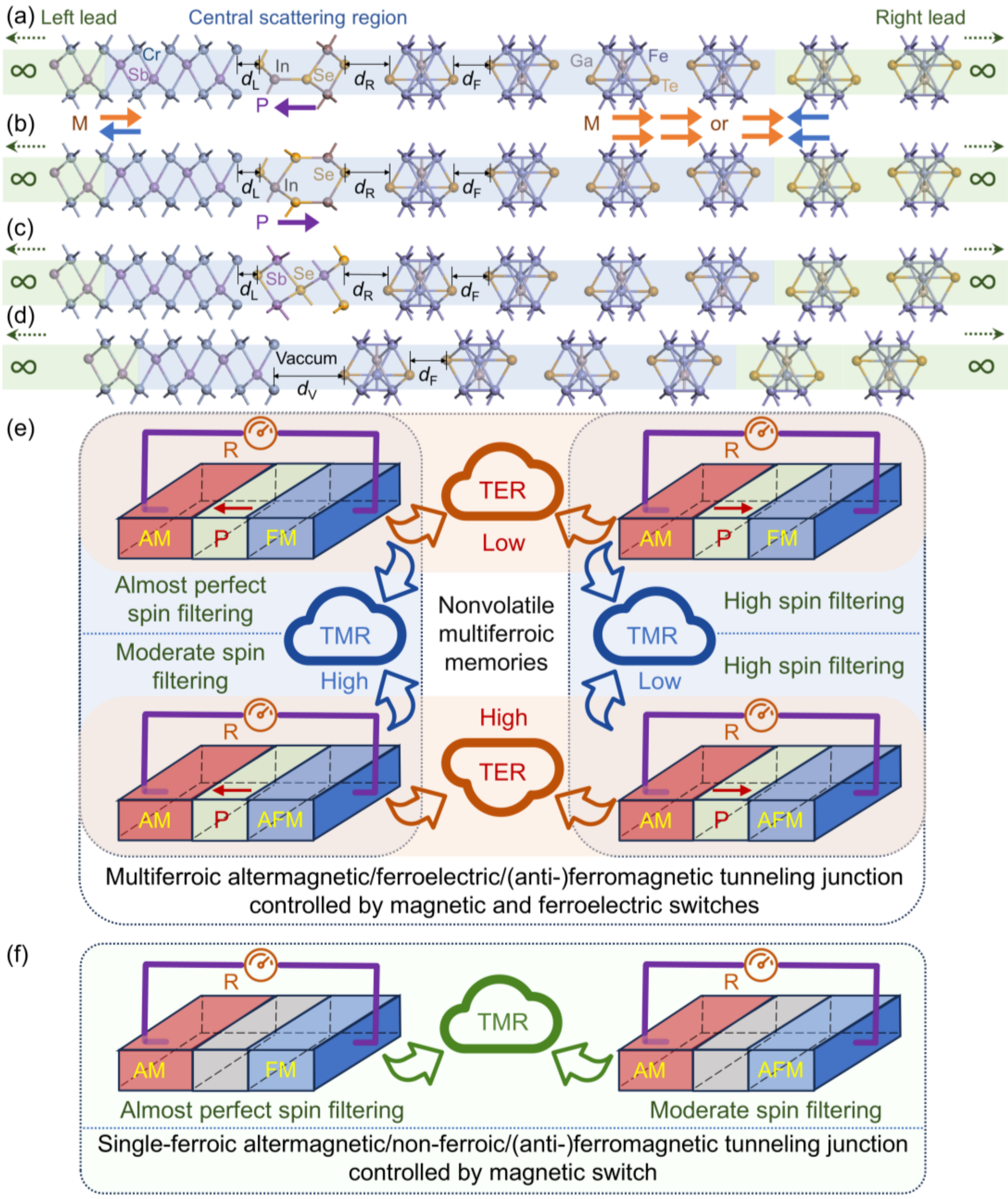

FIG. 2. Structural models of the proposed magnetic tunneling junctions with Cr interface: multiferroic $CrSb/In_2Se_3/Fe_3GaTe_2$ with different ferroelectric polarization directions (a,b); single-ferroic $CrSb/Sb_2Se_3/Fe_3GaTe_2$ (c), and $CrSb$/vacuum/$Fe_3GaTe_2$ (d). The optimized $d_F$ is 2.93 Å. The optimized interlayer distances $d_L/d_R$ are 1.80/3.58 Å in (a), 1.81/3.59 Å in (b), and 1.61/3.51 Å in (c), respectively. For the vacuum barrier (d), the separation distance $d_V$ is set to 5.50 Å in (d), approximately equal to the total thickness $d_L + d_R$ in (a-c). Schematic illustrations of the multiferroic (e) and single-ferroic (f) tunneling junctions demonstrate multiple resistance states and spin filtering effects that can reversibly tuned by magnetic and ferroelectric switches.

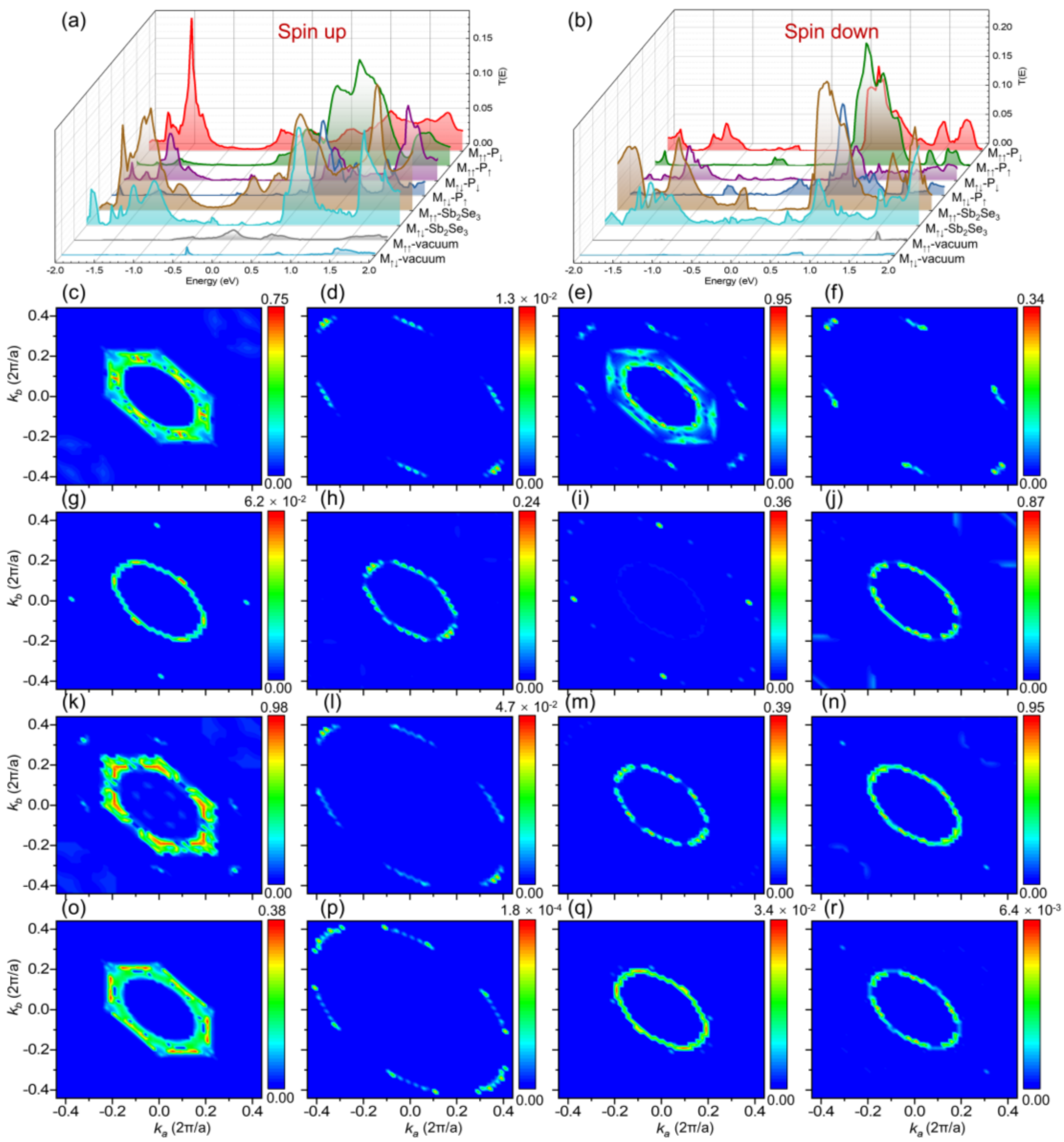

FIG. 3. Transmission coefficients as a function of energy in spin-up (a) and spin-down (b) channels. The $\vec{k}_{//}$-resolved transmission spectra in the 2D Brillouin zone for spin-up and spin-down channels of CrSb/$In_2Se_3$/$Fe_3GaTe_2$ with $M_{\uparrow\uparrow}$-$P_{\downarrow}$ (c,d), $M_{\uparrow\uparrow}$-$P_{\uparrow}$ (e,f), $M_{\uparrow\downarrow}$-$P_{\downarrow}$ (g,h) and $M_{\uparrow\downarrow}$-$P_{\uparrow}$ (i,j), CrSb/$Sb_2Se_3$/$Fe_3GaTe_2$ with $M_{\uparrow\uparrow}$ (k,l) and $M_{\uparrow\downarrow}$ (m,n), and CrSb/vacuum/$Fe_3GaTe_2$ with $M_{\uparrow\uparrow}$ (o,p) and $M_{\uparrow\downarrow}$ (q,r). With Cr interface, $M_{\uparrow\uparrow}$ and $M_{\uparrow\downarrow}$ stand for interlayer parallel and antiparallel magnetization of the $Fe_3GaTe_2$ electrodes, respectively. $P_{\downarrow}$ and $P_{\uparrow}$ are the opposite directions of ferroelectric polarization in $In_2Se_3$ barrier.

Table I. Spin-dependent electron transmission $T_\uparrow$ and $T_\downarrow$, TMR, TER, and spin filtering efficiency $\eta$ across CrSb/$In_2Se_3$/$Fe_3GaTe_2$ MTJs with Cr interface.

| | $M_{\uparrow\uparrow}$ | | | | $M_{\uparrow\downarrow}$ | | | | |
|---|---|---|---|---|---|---|---|---|---|
| | $T_\uparrow$ | $T_\downarrow$ | $T_{tot}$ | $\eta$ (%) | $T_\uparrow$ | $T_\downarrow$ | $T_{tot}$ | $\eta$ (%) | TMR (%) |
| $P_\downarrow$ | $2.89 \times 10^{-2}$ | $4.99 \times 10^{-5}$ | $2.89 \times 10^{-2}$ | ~100 | $6.99 \times 10^{-4}$ | $1.86 \times 10^{-3}$ | $2.56 \times 10^{-3}$ | 45 | 1031 |
| $P_\uparrow$ | $2.41 \times 10^{-2}$ | $1.31 \times 10^{-3}$ | $2.54 \times 10^{-2}$ | 90 | $7.72 \times 10^{-4}$ | $1.02 \times 10^{-2}$ | $1.10\times 10^{-2}$ | 86 | 132 |
| TER (%) | | 14 | | | | 328 | | | |

Table II. Spin-dependent electron transmission $T_\uparrow$ and $T_\downarrow$, TMR, and spin filtering efficiency $\eta$ across CrSb/$Sb_2Se_3$/$Fe_3GaTe_2$ and CrSb/vacuum/$Fe_3GaTe_2$ MTJs with Cr interface.

| | $M_{\uparrow\uparrow}$ | | | | $M_{\uparrow\downarrow}$ | | | | |
|---|---|---|---|---|---|---|---|---|---|
| Barrier | $T_\uparrow$ | $T_\downarrow$ | $T_{tot}$ | $\eta$ (%) | $T_\uparrow$ | $T_\downarrow$ | $T_{tot}$ | $\eta$ (%) | TMR (%) |
| $Sb_2Se_3$ | $4.66 \times 10^{-2}$ | $1.88 \times 10^{-4}$ | $4.68 \times 10^{-2}$ | ~100 | $3.01 \times 10^{-3}$ | $1.27 \times 10^{-2}$ | $1.57 \times 10^{-2}$ | 62 | 198 |
| Vacuum | $1.35 \times 10^{-2}$ | $1.20 \times 10^{-6}$ | $1.35 \times 10^{-2}$ | ~100 | $4.99 \times 10^{-4}$ | $6.27 \times 10^{-5}$ | $5.62 \times 10^{-4}$ | 78 | 2308 |

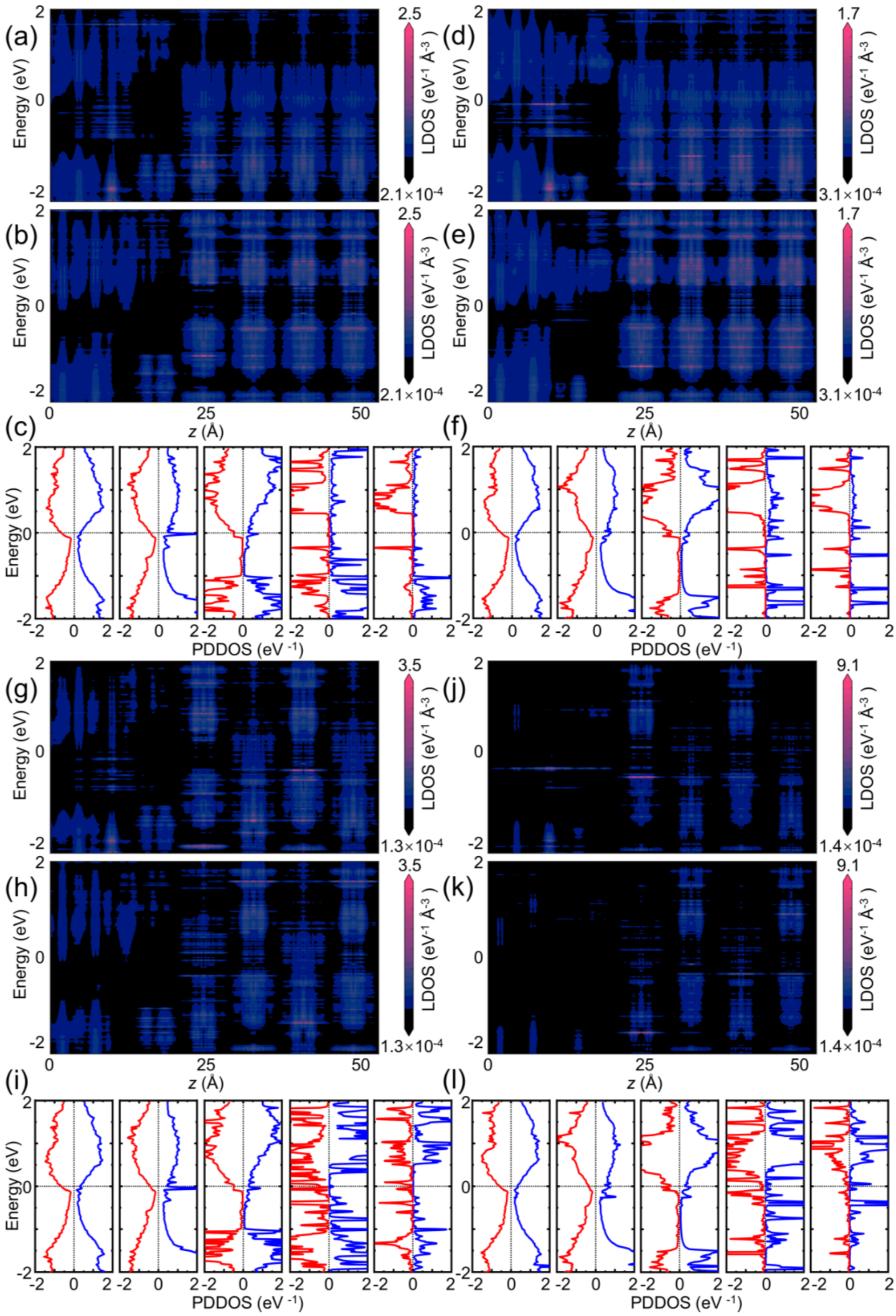

FIG. 4. Local density of states (LDOS) for spin-up and spin-down channels, and the spin- and layer-resolved projected device density of states (PDDOS) for $CrSb/In_2Se_3/Fe_3GaTe_2$ with Cr interface under different magnetic and ferroelectric configurations: $M_{\uparrow\uparrow}$-$P_{\downarrow}$ (a-c), $M_{\uparrow\uparrow}$-$P_{\uparrow}$ (d-f), $M_{\uparrow\downarrow}$-$P_{\downarrow}$ (g-i) and $M_{\uparrow\downarrow}$-$P_{\uparrow}$ (j-l). In the PDDOS plots, the blue and red lines stand for the spin-up and spin-down states, respectively. The electronic contributions originate from the left CrSb electrodes serving as the source.

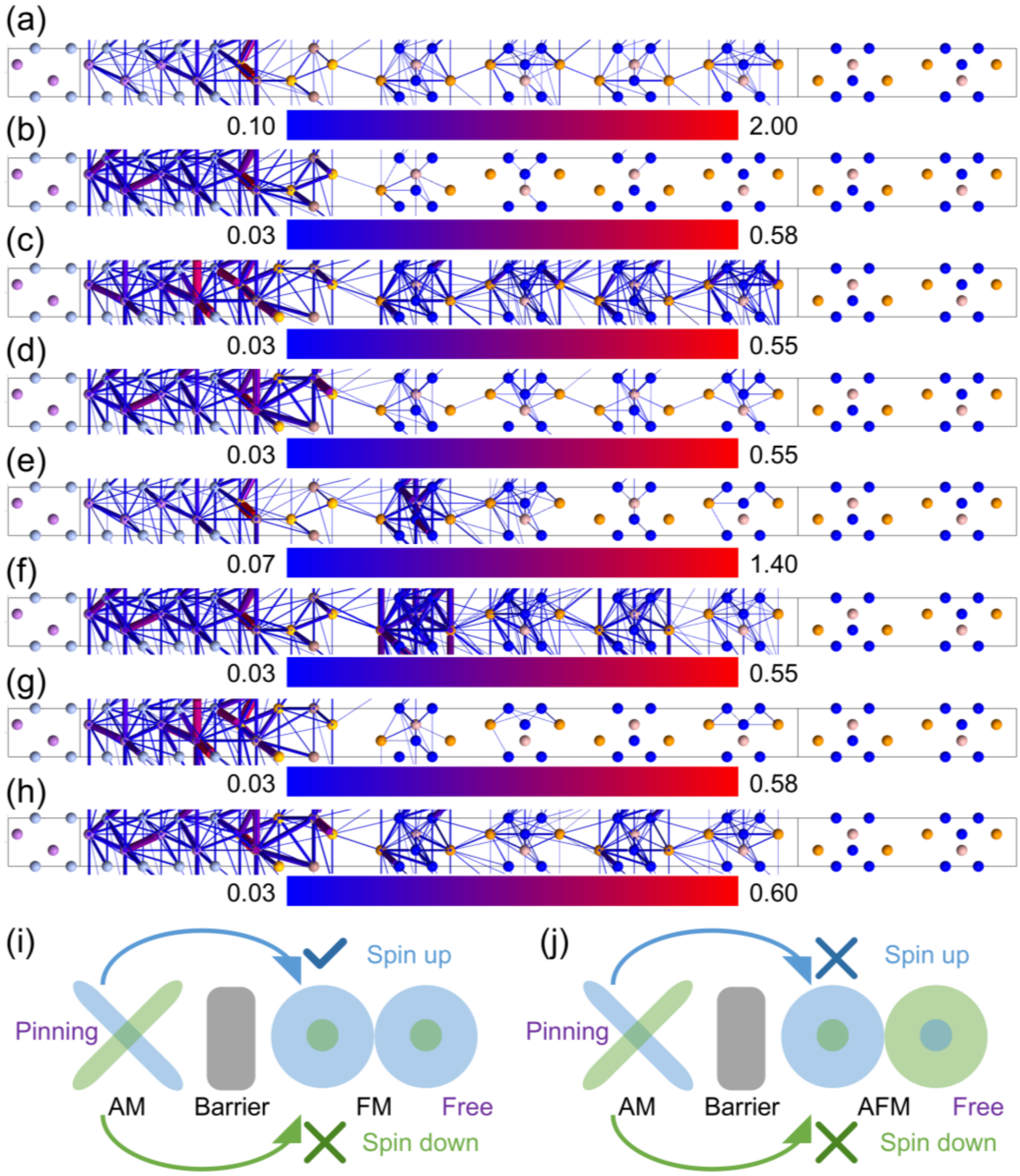

FIG. 5. Transmission pathways in spin-up and spin-down channels for $CrSb/In_2Se_3/Fe_3GaTe_2$ with Cr interface under $M_{\uparrow\uparrow}$-$P_{\downarrow}$ (a,b), $M_{\uparrow\uparrow}$-$P_{\uparrow}$ (c,d), $M_{\uparrow\downarrow}$-$P_{\downarrow}$ (e,f) and $M_{\uparrow\downarrow}$-$P_{\uparrow}$ (g,h), in which the source of contribution is set as the left CrSb electrodes. The schematics for ferroic tunnel junctions of altermagnetic electrode/non-magnetic barrier/(anti)ferromagnetic electrode with parallel (i) and antiparallel (j) spin configurations, where altermagnet (AM) is the pinning layer and ferromagnet/antiferromagnet (FM/AFM) is the free layer.

Table III. Spin-dependent electron transmission $T_\uparrow$ and $T_\downarrow$, TMR, TER, and spin filtering efficiency $\eta$ across $CrSb/In_2Se_3/Fe_3GaTe_2$ MTJs with Sb interface.

| | $M_{\uparrow\uparrow}$ | | | | $M_{\uparrow\downarrow}$ | | | | |
|---|---|---|---|---|---|---|---|---|---|
| | $T_\uparrow$ | $T_\downarrow$ | $T_{tot}$ | $\eta$ (%) | $T_\uparrow$ | $T_\downarrow$ | $T_{tot}$ | $\eta$ (%) | TMR (%) |
| $P_\downarrow$ | $1.18 \times 10^{-2}$ | $9.17 \times 10^{-5}$ | $1.19 \times 10^{-2}$ | ~100 | $1.83 \times 10^{-4}$ | $8.44 \times 10^{-4}$ | $1.03 \times 10^{-3}$ | 64 | 1056 |
| $P_\uparrow$ | $1.93 \times 10^{-2}$ | $2.15 \times 10^{-4}$ | $1.95 \times 10^{-2}$ | 98 | $3.31 \times 10^{-4}$ | $7.96 \times 10^{-3}$ | $8.29 \times 10^{-3}$ | 92 | 136 |
| TER (%) | | 65 | | | | 707 | | | |

Table IV. Spin-dependent electron transmission $T_\uparrow$ and $T_\downarrow$, TMR, and spin filtering efficiency $\eta$ across $CrSb/Sb_2Se_3/Fe_3GaTe_2$ and $CrSb/vacuum/Fe_3GaTe_2$ MTJs with Sb interface.

| | $M_{\uparrow\uparrow}$ | | | | $M_{\uparrow\downarrow}$ | | | | |
|---|---|---|---|---|---|---|---|---|---|
| Barrier | $T_\uparrow$ | $T_\downarrow$ | $T_{tot}$ | $\eta$ (%) | $T_\uparrow$ | $T_\downarrow$ | $T_{tot}$ | $\eta$ (%) | TMR (%) |
| $Sb_2Se_3$ | $4.79 \times 10^{-2}$ | $4.51 \times 10^{-4}$ | $4.84 \times 10^{-2}$ | ~100 | $8.12 \times 10^{-4}$ | $1.06 \times 10^{-2}$ | $1.14 \times 10^{-2}$ | 86 | 325 |
| Vacuum | $5.28 \times 10^{-3}$ | $1.11 \times 10^{-4}$ | $5.39 \times 10^{-3}$ | 96 | $2.40 \times 10^{-4}$ | $7.44 \times 10^{-4}$ | $9.84 \times 10^{-4}$ | 51 | 448 |

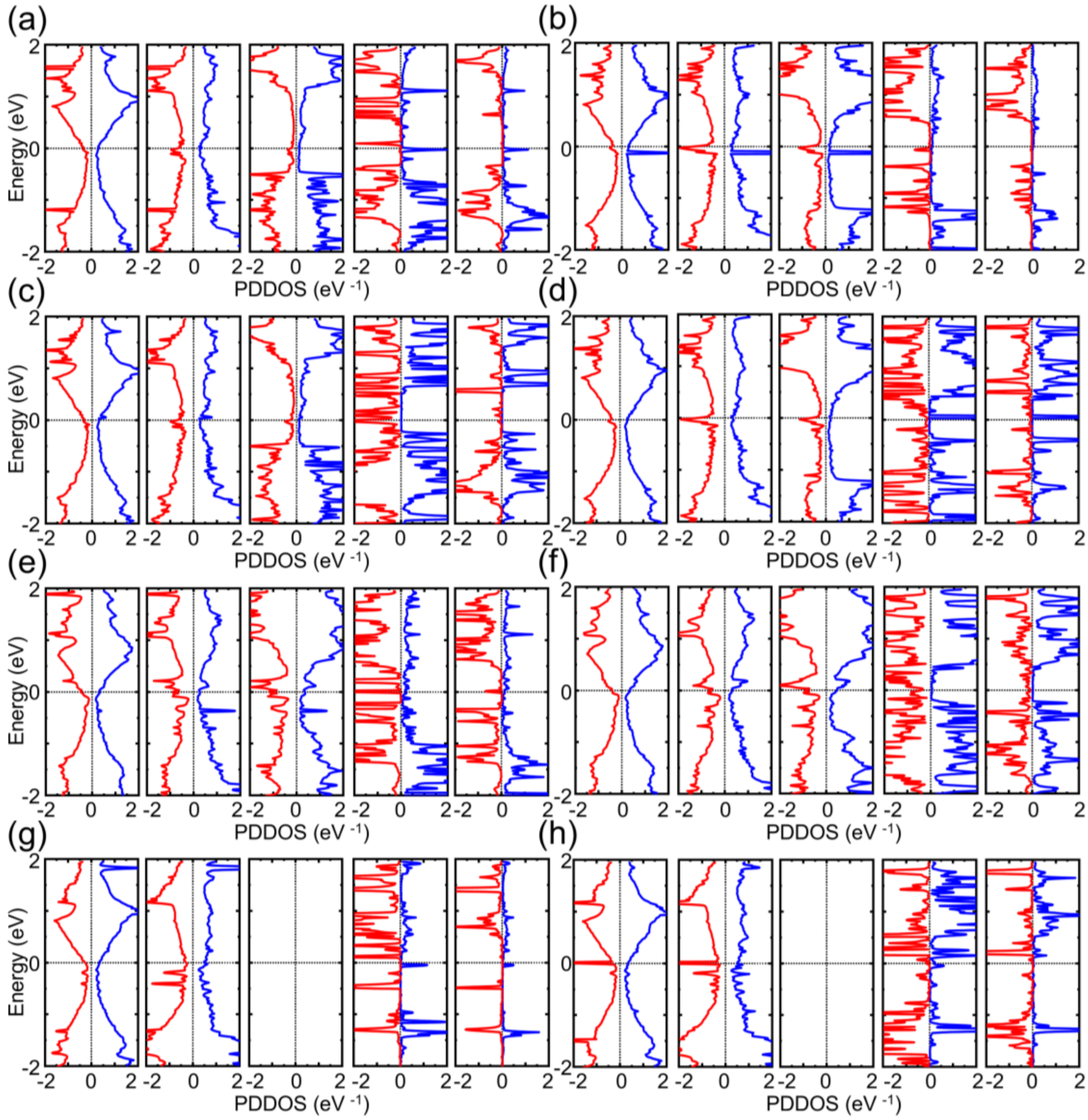

FIG. 6. The spin- and layer-resolved projected device density of states (PDDOS) under different magnetic and ferroelectric configurations with Sb interface. For CrSb/$In_2Se_3$/$Fe_3GaTe_2$: $M_{\uparrow\uparrow}$-$P_{\downarrow}$ (a), $M_{\uparrow\uparrow}$-$P_{\uparrow}$ (b), $M_{\uparrow\downarrow}$-$P_{\downarrow}$ (c), and $M_{\uparrow\downarrow}$-$P_{\uparrow}$ (d). For CrSb/$Sb_2Se_3$/$Fe_3GaTe_2$: $M_{\uparrow\uparrow}$ (e), and $M_{\uparrow\downarrow}$ (f). For CrSb/vacuum/$Fe_3GaTe_2$: $M_{\uparrow\uparrow}$ (g), and $M_{\uparrow\downarrow}$ (h).

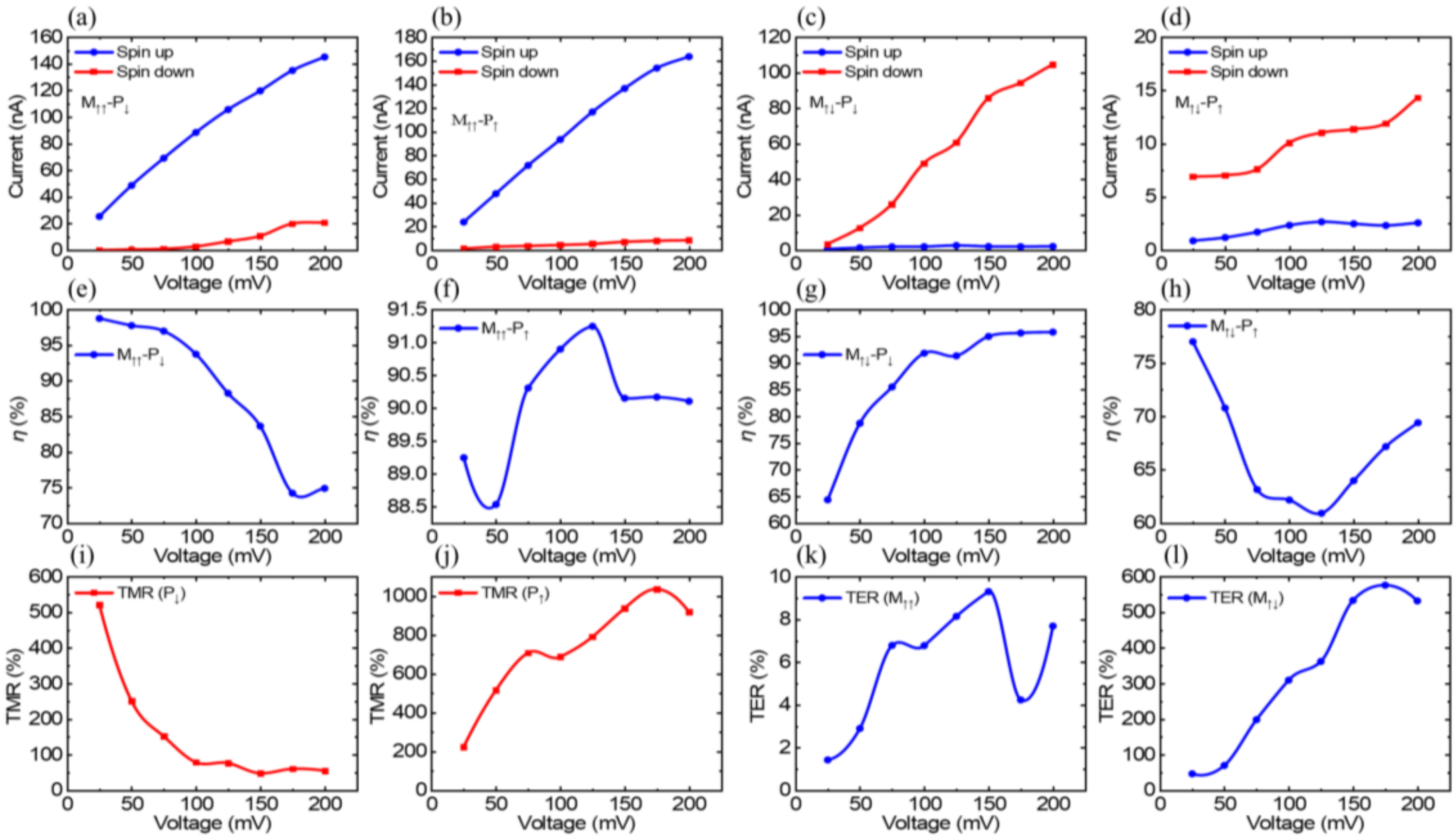

FIG. 7. Spin-dependent current (a-d), spin filtering efficiency (e-h), tunneling magnetoresistance (i,j), and tunneling electroresistance (k,l) across $CrSb/In_2Se_3/Fe_3GaTe_2$ MTJ with Cr interface under bias voltage.